\documentclass[aps,prd,letterpaper,preprintnumbers,floatfix,superscriptaddress,nofootinbib,onecolumn]{revtex4}
\pdfoutput=1 
\usepackage{graphicx,dcolumn,bm,epsfig,cancel,hyperref}
\usepackage{multirow}
\usepackage{amsmath}
\usepackage{amssymb, times}
\usepackage[utf8]{inputenc}
\usepackage{color}

\usepackage{slashed}
\usepackage{soul}
\hypersetup{colorlinks=true,
	breaklinks=true,
    linkcolor=blue,          
    citecolor=blue,        
    urlcolor=blue           
}

\begin{document}
\title{
The 7\% Rule: A Maximum Entropy Prediction on New Decays of the Higgs Boson 
}
\author{Alexandre Alves}
\email{aalves@unifesp.br}
\affiliation{Departamento de Física, Universidade Federal de São Paulo, UNIFESP, Diadema-SP, Brazil}

\author{Alex Gomes Dias}
\email{alex.dias@ufabc.edu.br}
\affiliation{Centro de Ci\^encias Naturais e Humanas, Universidade Federal do ABC, UFABC, Santo Andr\'e-SP, Brazil,}

\author{Roberto da Silva}
\email{rdasilva@if.ufrgs.br}
\affiliation{Instituto de Física, Universidade Federal do Rio Grande do Sul, UFRGS, Porto Alegre-RS, Brazil}

\begin{abstract}
    The entropy of the Higgs boson decay probabilities distribution in the Standard Model (SM) is maximized for a Higgs mass value that is less than one standard deviation away from the current experimental measurement. This successful estimate of the Higgs mass encourages us to propose tests of the Maximum Entropy Principle (MEP) as a 
    tool for theoretical inferences in other instances of Higgs physics.
    In this letter, we show that, irrespective of the extension of the SM predicting a new Higgs boson decay channel, its branching ratio can be inferred to be around 7\% in such a way that the new entropy of decays still exhibits a maximum at the experimental Higgs mass. This 7\% rule can be tested whenever a new Higgs decay channel is found. In order to illustrate the MEP predictions, we apply the MEP inference to 
    Higgs portal models, Higgs-axion interactions, lepton flavour violating decays of the Higgs boson, and a dark gauge boson model. 
\end{abstract}


\maketitle

\section{Introduction}
\label{sec:Intro}

 Naturalness and the hierarchy problem of the Standard Model (SM) have been some of the leading guidelines in the quest for a theory beyond the Standard Model (BSM). It boils down to the difficulty to understanding
 why the Higgs boson mass is of the electroweak energy scale once loop contributions are large ~\cite{Senjanovic:2020pnq,Dvali:2019mhn,Dijkstra2019NaturalnessAA,Williams:2015gxa}. 
 Protecting the Higgs mass from receiving those large contributions can be accomplished by postulating new symmetries of nature like the fermion $\leftrightarrow$ boson symmetry of supersymmetric models~\cite{Martin:1997ns}, for example. Another solution is bringing the Planck scale down closer to the electroweak scale, as proposed in extra dimension models~\cite{Cheng:2002ab}. If the Higgs boson is not a fundamental particle, but a bound state made of quarks tied together by a new confining force is yet another solution to the problem~\cite{Chivukula:1998wd,Hill:2002ap}. Whatever is the solution, however, new particles and interactions are common features of models and mechanisms to solve the long standing problem of the SM Higgs boson mass, otherwise the contributions to the Higgs mass seem to be very finely tuned in order it lies at the electroweak scale. 
 
 Fine tuning problems as the one behind the Higgs mass can also be found in another long standing puzzle -- the smallness of the cosmological constant. 
 Just like the Higgs boson mass, the cosmological constant also poses an enormous problem to the standard cosmological theory -- its current value is around a hundred orders of magnitude smaller than that expected from field theory computations including loop corrections. That is it, the cosmological constant seems to be rather unnatural and extremely fine tuned as well. Solving these problems will lead to new fundamental understanding of nature, however testing the proposed solutions might be difficult in the near future, notably in respect the quantum aspect of gravity.

 On the other hand, applications of entropic principles as inference tools, in particular, the Maximum Entropy Principle (MEP)~\cite{PhysRev.106.620}, have found great success across all sciences, including physics. For example, MEP and the Causal Entropic Principle were able to determine quite accurately the Higgs boson mass~\cite{Alves:2014ksa} and the cosmological constant~\cite{Bousso:2007kq}, respectively, without assuming new physics beyond the standard theories. Whether there exists a thermodynamic mechanism at work that fixes these parameters or not is a theoretical possibility that can be further investigated. The existence of such mechanism, however, does not exclude fundamental new physics by any means, once new particles and interactions are also bound to the thermodynamic principles. Whatever is the case, entropic principles have proven their usefulness as theoretical tools for statistical inference and prediction and establishing their correctness and accuracy can shed some light in those fundamental problems. In fact, Jaynes sustained that statistical mechanics is a kind of statistical inference tool, unifying the concept of entropy from information theory with the Boltzmann definition, and derived key thermodynamics equations from MEP~\cite{PhysRev.106.620}. The force of MEP inference, just like using statistical mechanics methods, is that it is possible to get useful information about a system without knowing its fine details.
 
 The application of MEP in particle physics has been an interesting line of research. It was first observed in Ref.~\cite{dEnterria:2012eip} that the measured Higgs mass is very close to the maximum likelihood estimation based on the Higgs branching ratios in the SM. Subsequently, we showed in Ref.~\cite{Alves:2014ksa} that the Higgs boson mass can be precisely inferred through MEP, with the Gibbs-Shannon entropy function built from the Higgs decay probabilities. Apart from the successful application to the Higgs mass inference, we have also used MEP inference in effective models with axion-like particles~\cite{Alves:2017ljt}. 
 The use of the Gibbs-Shannon entropy of the distribution of the branching ratios to look for new channels in hadron physics has been presented in Refs.~\cite{Millan:2018fme,Llanes-Estrada:2017clj}. Our approach to the same problem in Higgs physics is different, however. A number of other interesting applications of MEP to particle physics has also appeared in the literature, see for example Refs.~\cite{Karapetyan:2017edu,Karapetyan:2018oye,Karapetyan:2019ran,Braga:2018fyc,Ma:2018wtw,daSilva:2017jay,Braga:2020myi,Ferreira:2020iry}. 
 For an introduction to MEP and information theory, see Refs.~\cite{Hanel6905,Witten:2018zva,10.5555/1146355}, for example.
 
 
 In this work, we show that insisting that a new decay channel of the Higgs boson still respects MEP, in the sense that the addition of the new decay channel does not spoil the agreement between the experimental and MEP-inferred masses, leads to a model independent prediction on the branching ratio of the new model: it must be around 7\% of the total Higgs decays, this is {\it the 7\% rule}.  This prediction can be tested experimentally. If MEP predicts that a single new decay channel of the Higgs boson is not compatible with the experimental Higgs mass, but evidence for it is found, then MEP can be falsified. On the other hand, finding a new decay channel of the Higgs boson with parameters correctly inferred by MEP adds evidence to its correctness. Measuring the Higgs boson width is another way to test this prediction. Recently, the CMS Collaboration measured the Higgs width as $3.2^{+2.8}_{-2.2}$ MeV~\cite{Sirunyan:2019twz} which is already close to the SM value of 4.07 MeV. In the next run of the LHC, the uncertainty of the measurement will shrink and we might test this universal MEP prediction of a single new channel with 7\% of branching fraction. If MEP can be put to test in order to establish it as an accurate inference tool, as the example of the Higgs boson mass suggests, it might become useful for phenomenological studies in particle physics. The benefits of having a functional tool that spots the right parameters of a model would be immense.
 
 The prediction that a new channel should occur 7\% of the time is intuitive as we are going to discuss, but an inference tool should also be able to give an estimate of the confidence belt. For that goal, it is necessary to obtain the correct dependence of the new branching ratio in terms of the various parameters of the Standard Model and also of those of the new model as predicted by MEP. We work out that inference tool by computing the solution to a differential equation which expresses the Maximum Entropy Principle. 
 
 We then applied the MEP inference, taking into account experimental uncertainties in the SM parameters, to four BSM scenarios predicting a new Higgs decay channel in order to illustrate the inference, namely, (1) a class of Higgs-portal models with fermionic, scalar and vector dark matter, (2) a lepton flavor violation model, (3) a dark gauge boson coupling to the Higgs boson, and (4) a Higgs into an axion like particle pair decay. We found that the MEP inference is able to sharply spot the parameters of the new models where, according to the principle, a new Higgs decay channel should manifest itself. 
 
 Our paper is organized as follows. In Section~\ref{sec:BSMEP} we compute the entropy of the Higgs boson decay probabilities with a new decay channel beyond those of the SM; in Section~\ref{sec:MEPH}, we obtain the branching ratio of the new decay channel from MEP; in Section~\ref{sec:typeI} we apply the inference to new physics models; finally, we present our conclusion in Section~\ref{sec:conclusions}.
 

\section{The entropy of the Higgs boson decays with a new channel}
\label{sec:BSMEP}

Let us review the basic steps of the MEP inference of the Higgs boson mass, $m_h$~\cite{Alves:2014ksa}. Consider an ensemble of $N$ non-interacting Higgs bosons that are allowed to decay into the $M$ basic 2-body SM channels: $\gamma\gamma,\; gg,\; Z\gamma,\; ZZ,\; WW,\; q\bar{q},\; \ell^+\ell^-$ plus a new one $\chi\chi$ as long as $m_\chi < m_h/2$, where $m_\chi$ is the mass of the particle $\chi$. We consider Higgs masses such that decays for all the quark flavors $q=u,d,s,c,b$ but the top quark and all the leptons flavors are also included $\ell=e,\mu,\tau$. It turns out to be that $M=13$ but we will keep $M$ in the following formulae for the generality's sake.

The probability of a given configuration of the $N$ Higgs bosons, after their decays, into the $M+1=14$ final states listed above, is given by a multinomial distribution 
\begin{equation}
P(\left[n_k\right]_{k=1}^{M+1})= \frac{N!}{n_1!\cdots\ n_M!}\prod_{k=1}^{M+1} BR^{n_k}_k(m_h)\; ,
\end{equation}
where $n_{\gamma\gamma},\cdots, n_{\chi\chi}$ are the occupation numbers of each final state, and the branching ratios 
\begin{equation}
BR_i(m_h)=\frac{\Gamma_i}{\sum_{i=1}^{M+1}\Gamma_i}=\frac{\Gamma_i}{\Gamma_{SM}+\Gamma_\chi},\;\; i=1,\cdots, M+1\; 
\end{equation}
are calculated from the total width of the SM, $\Gamma_{SM}=\sum_{i=1}^{M}\Gamma^{SM}_i$ with $\Gamma^{SM}_i=\Gamma_i, i=1,\cdots, M$ the partial widths of the Higgs decay channels and $\Gamma_\chi$ is the partial width of the new channel,
identified as the 14th channel. The normalization property of the branching ratios is $BR_\chi+\sum_{i=1}^{M} BR^{SM}_i = 1$. From now on, we denote the Higgs branching ratio into SM channels as $BR_i\equiv BR^{SM}_i,i=1,\cdots, M$, and the new one as $BR_{M+1}=BR_\chi$. The partial widths and branching ratios of the SM channels and of the new channel depend on other parameters which are not being explicitly shown at this stage.

The Gibbs-Shannon entropy of the $N$ Higgs decays is given by
\begin{equation}
    S_N =  -\sum_{\{n\}}^N
    P(\left[n_k\right]_{k=1}^{M+1}) \ln[P(\left[n_k\right]_{k=1}^{M+1})]=-\langle \ln P
    \rangle\; ,
\end{equation}
where $\sum_{\{n\}}^N(\bullet)=\sum_{n_1=0}^N\cdots\sum_{n_{M+1}=0}^N(\bullet)\delta\left(N-\sum_{i=1}^{M+1}n_i\right)$.
An asymptotic formula for this sum has been derived in Ref.~\cite{Cichon} up to terms of ${\cal O}(1/N)$
\begin{eqnarray}
S_N &=& \frac{M}{2}\ln(2\pi N)+\frac{1}{2}\ln\left(\prod_{i=1}^{M+1} BR_i\right)\nonumber\\
&+& \frac{1}{12N}\left(3M+1-\sum_{i=1}^{M+1}\frac{1}{BR_i}\right)+{\cal O}\left(\frac{1}{N^2}\right)\; .
\end{eqnarray}

In the limit where $N$ is very large, we can drop the terms suppressed by $1/N$, whereas the term $\ln\left(\prod_{i=1}^{M+1} BR_i\right)$ embodies all the dependence on the Higgs dynamics. The term $\ln(2\pi N)$ is just a large constant for our aims once we are interested only in the variation of $S_N$ with $m_h$. We then define the term which varies with the parameters of the model
\begin{equation}
    S_{BSM} = \frac{1}{2}\ln\left(\prod_{i=1}^{M+1} BR_i\right)=\frac{1}{2}\ln BR_\chi+\sum_{i=1}^{M}\frac{1}{2}\ln BR_i^{SM}\; ,
    \label{eq:sbsm}
\end{equation}
this is the entropy of the beyond the Standard Model Higgs boson decay probabilities, including the new decay channel.

Now let us also define the following ratios
\begin{equation}
    F_\chi=\frac{\Gamma_\chi}{\Gamma_{SM}},\;\; F_i^{SM}=\frac{\Gamma^{SM}_i}{\Gamma_{SM}}\; ,
    \label{eq:brratios}
\end{equation}
in terms of which the branching ratios can be written as follows
\begin{equation}
    BR_\chi=\frac{F_\chi}{1+F_\chi},\;\; BR_i^{SM}=\frac{F^{SM}_i}{1+F_\chi}, i=1,\cdots, M\; .
    \label{eq:BRratios}
\end{equation}

Now, it is easy to show that
\begin{eqnarray}
    S_{BSM} &=& S_{SM}+\frac{1}{2}\ln\frac{F_\chi}{(1+F_\chi)^{M+1}}\label{eq:Ssbm}, \\
    S_{SM} &=& \sum_{i=1}^{M}\frac{1}{2}\ln F_i^{SM}\label{eq:Ssm}, \;
\end{eqnarray}
where $S_{SM}$ is what we call the entropy of the Higgs decay probabilities distribution in the Standard Model. The Eqs.~\eqref{eq:sbsm} and \eqref{eq:Ssbm} above reflect the additive property of the entropy.

\section{The Maximum Entropy Principle in Higgs decays}
\label{sec:MEPH}
 The Maximum Entropy Principle as applied to the SM Higgs boson mass determination can be expressed concisely as
 \begin{equation}
      m_h^{sm} = \underset{m_h}{\mathrm{argmax}}\; S_{SM}(m_h,\boldsymbol{\theta}_{SM}),
 \end{equation}
 where $\boldsymbol{\theta}_{SM}$ represents the parameters of the SM which the branching ratios depend upon and $S_{SM}$ given by Eq.~\eqref{eq:Ssm}. This is nothing but the Maximum Likelihood Estimation (MLE) of the Higgs mass based on the decay probabilities of the Higgs boson in the SM and that was applied in Ref.~\cite{dEnterria:2012eip}. In MEP~\cite{PhysRev.106.620}, the entropy is a means for statistical inference on multinomial process, as the one where an ensemble of Higgs bosons decay to a number of possible channels with parametrized probabilities given by the branching ratios, and leads as shown in the previous section, to the MLE estimate.
 
\begin{figure}
     \centering
     \includegraphics[scale=0.6]{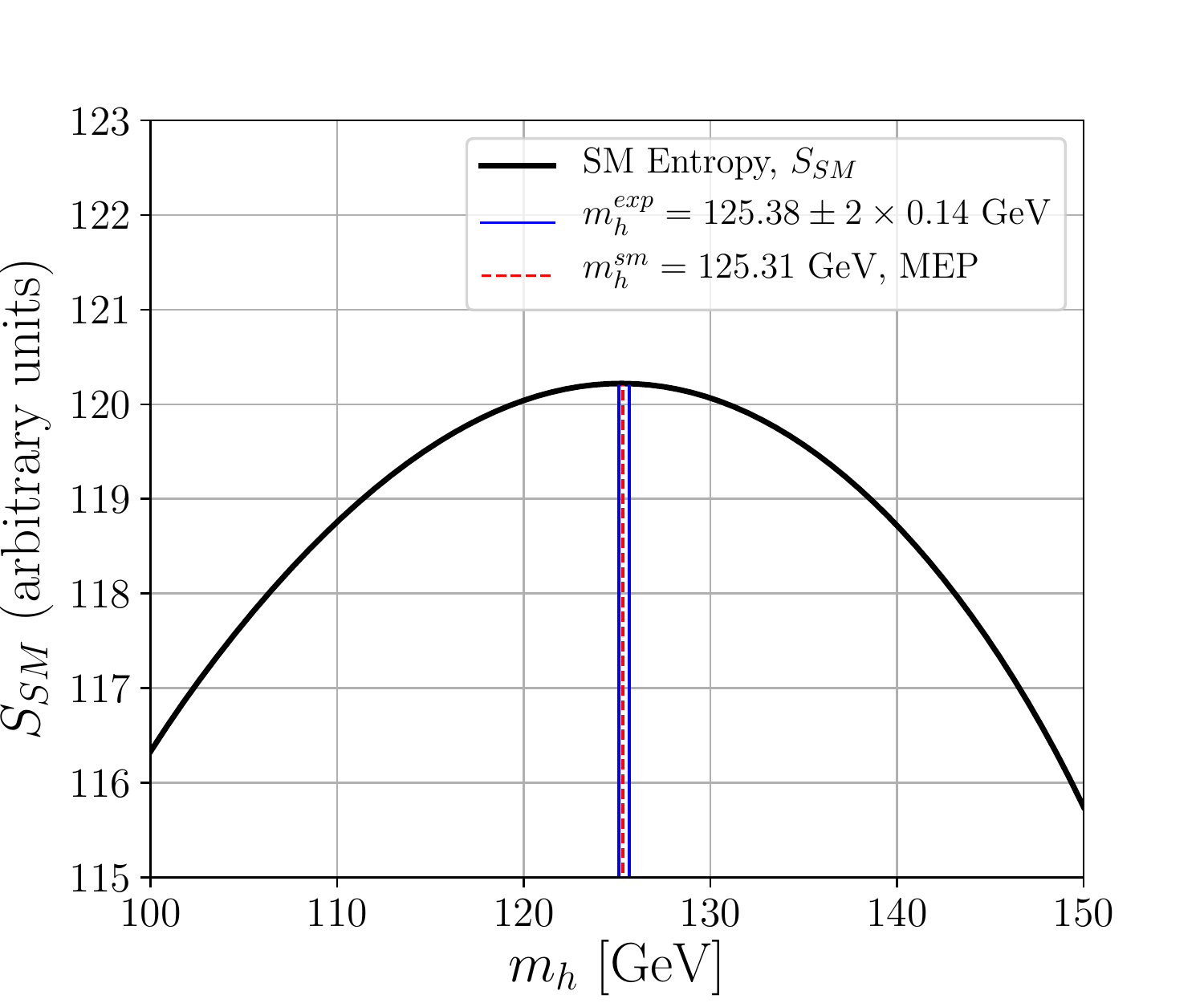}
     \caption{Entropy of the SM Higgs boson decays. The vertical red line indicates the mass that maximizes the entropy while the blue vertical lines bound the most precise $2\sigma$ experimental measurement of the Higgs mass to date~\cite{Sirunyan:2020xwk}.}
     \label{fig:Ssm}
 \end{figure}

 In Figure~\eqref{fig:Ssm}, we show $S_{SM}$ as a function of $m_h$ (in GeV).
 The dashed red line indicates the location of the maximum of $S_{SM}$, this is the SM Higgs mass inferred from MEP,
 \begin{equation}
 m_h^{sm}=125.31\pm 0.20\; \hbox{GeV}, 
 \label{eq:mhat}
\end{equation} 
 and the solid blue lines, the 95\% CL region of experimental value from the most precise CMS measurement~\cite{Sirunyan:2020xwk}
 \begin{equation}
     m_h^{exp}=125.38\pm 0.14\; \hbox{GeV}. 
     \label{eq:mexp}
 \end{equation}

 The central value obtained from MEP, with SM parameters taken from the PDG~\cite{PhysRevD.98.030001}, is less than one standard deviation away from the CMS value whose precision reaches a milestone of 0.1\%. All the branching ratios of the Higgs boson in the SM are calculated using an adapted version of the \texttt{HDECAY}~\cite{Djouadi:2018xqq} package. The uncertainty in the MEP inferred mass of Eq.~\eqref{eq:mhat} is obtained by marginalizing over the uncertainties of the all the SM parameters which affect the calculation of the branching ratios. 
 
 The astonishing agreement between the MEP inference and the experimental value, which has converged to the MEP prediction since the first Higgs mass measurements~\cite{Alves:2014ksa}, encourages us to extend the Maximum Entropy Principle to models beyond the SM. In Ref.~\cite{Alves:2017ljt}, MEP is applied for inferences of an axion-like particle decaying to photons and neutrinos in search for clues for the parameters of the model. This time, we are interested in deriving a general tool for BSM  inferences involving the 125 GeV Higgs boson of a model predicting a new decay channel. It has to be emphasized that we are considering only one new decay channel. The case where we have two or more decay channels open can be treated more easily by numerical means as done in Ref.~\cite{Alves:2017ljt}.
 
The MEP inference tool applied to the BSM Higgs entropy
\begin{equation}
    \frac{d}{dm_h} S_{BSM}\Bigg|_{m_h=\hat{m}_h}= 0\; 
    \label{eq:mepBSM}
\end{equation}
must hold for the true Higgs boson mass, $\hat{m}_h$. This is the fundamental equation from which the parameters of any new physics model predicting a new Higgs decay channel are inferred by the Maximum Entropy Principle. The model independent aspect of our proposal to extend the application of MEP beyond the Standard Model is contained in this equation.


Substituting Eq.~\eqref{eq:Ssbm} into Eq.~\eqref{eq:mepBSM}, we find
\begin{equation}
  \left(\frac{1}{F_\chi}-\frac{M+1}{1+F_\chi}\right)\frac{d F_\chi}{dm_h}\Bigg|_{m_h=\hat{m}_h} = -\frac{dS_{SM}}{dm_h}\Bigg|_{m_h=\hat{m}_h}\; .
  \label{eq:edo1}
\end{equation}

 Assuming that the true Higgs mass should be very close to the experimental Higgs mass, and observing that the experimental mass is very close to the mass the maximizes $S_{SM}$, that is it, $\hat{m}_h\approx m_h^{exp}\approx m_h^{sm}$, we can affirm that, to a very good approximation, $dS_{SM}/dm_h\approx 0$ at $m_h=\hat{m}_h$. Then, approximately, Eq.~\eqref{eq:edo1} reads
\begin{equation}
  \left(\frac{1}{F_\chi}-\frac{M+1}{1+F_\chi}\right)\Bigg|_{m_h=\hat{m}_h} \times \frac{d F_\chi}{dm_h}\Bigg|_{m_h=\hat{m}_h}\approx 0\; .
  \label{eq:edo2}
\end{equation}

 If $F_\chi$ is not itself maximized by the true mass, then $F_\chi(\hat{m}_h)\approx \frac{1}{M}$. Actually, we can also obtain that
 \begin{equation}
    F_\chi(m_h^{exp})\approx F_\chi(m_h^{sm})\approx\frac{1}{M}\Longrightarrow BR_\chi(m_h^{exp})\approx BR_\chi(m_h^{sm})\approx \frac{1}{M+1}=7.1\%\; \hbox{if}\; M=13\; .
    \label{eq:edo3}
\end{equation}
This is the 7\% rule predicted by MEP for a new Higgs decay channel, irrespective of the type of new particle and its interaction with the 125 GeV Higgs boson.

Treating $F_\chi(\hat{m}_h)$ as our unknown, Eq.~\eqref{eq:edo1} is an ordinary differential equation which can be easily integrated to give
\begin{equation}
    \ln\frac{F_\chi}{(1+F_\chi)^{M+1}}=-S_{SM}+C\; .
\end{equation}
In this work we are interested in a closed form solution for the new branching ratio. We therefore need a boundary condition to fix $C$. As we now, $F_\chi(m_h^{sm})\approx 1/M$ to a very good precision. If we thus fix $F_\chi(m_h^{sm})=1/M$, we can say, again to a good approximation, that
\begin{equation}
    \ln\frac{F_\chi}{(1+F_\chi)^{M+1}}\Bigg|_{m_h=m_h^{sm}} \approx -S_{SM}\Bigg|_{m_h=m_h^{sm}} + C\Longrightarrow C\approx S_{SM}\Bigg|_{m_h=m_h^{sm}}+\ln\frac{M^M}{(M+1)^{M+1}}\; ,
\end{equation}
and the implicit solution for the new branching ration is given by

\begin{equation}
    BR_\chi(1-BR_\chi)^M\approx \frac{M^M}{(M+1)^{M+1}}e^{-\Delta S_{SM}(\hat{m}_h,\boldsymbol{\theta}_{exp}^{SM})},
    \label{eq:sol}
\end{equation}
where $\boldsymbol{\theta}_{exp}^{SM}$ denotes the current experimental values of the SM parameters, and by using Eq.~\eqref{eq:BRratios} and defining 
\begin{equation}
    \Delta S_{SM}(\hat{m}_h,\boldsymbol{\theta}_{exp}^{SM})\equiv S_{SM}(\hat{m}_h,\boldsymbol{\theta}_{exp}^{SM})-S_{SM}(m^{sm}_h,\boldsymbol{\theta}_{exp}^{SM})\; .
    \label{eq:deltaS}
\end{equation}

The best estimate for the true Higgs mass is the current experimental mass of Eq.~\eqref{eq:mexp}. Assuming the current SM parameters, $\Delta S_{SM}(m_h^{exp},\boldsymbol{\theta}_{exp}^{SM})\approx -1.4\times 10^{-4}$. The mismatch between $S_{SM}(m_h^{exp})$ and $S_{SM}(m^{sm}_h)$ is due the uncertainties in the SM parameters and possibly a missing channel. 

The recipe of the proposed MEP inference for the central value of the new Higgs branching ratio can be stated as follows
\begin{equation}
    BR_\chi(m_h^{exp},\boldsymbol{\theta}_{exp}^{SM},\boldsymbol{\theta}_\chi) = \frac{1}{14}\approx 7.1\%\;\; ,
    \label{eq:br_central}
\end{equation}
where $\boldsymbol{\theta}_\chi$ denote the parameters of the new model that the branching ratio of $\chi$ depends upon.

It is interesting to note that adding a new decay channel to the Higgs boson for the case where $m^{exp}_h=m^{sm}_h$ corresponds precisely to assume the less informative guess for the new probability function. If the only constraint on the Higgs branching ratio was $\sum_{i=1}^{M+1} BR_i=1$, the MEP inference of the branching ratios would be $BR_i=\frac{1}{M+1},\; i=1,\cdots, M+1$, so it seems to be consistent that the new channel is predicted to have a probability of decay of $\sim \frac{1}{M+1}$ if it is supposed to maximize the entropy of decays.


If $\Delta S_{SM}(m_h^{exp},\boldsymbol{\theta}_{exp}^{SM})<0$, there is no real solution to Eq.~\eqref{eq:sol}. Uncertainties in the experimental Higgs mass and the SM parameters can change the sign of  $\Delta S_{SM}(m_h^{exp},\boldsymbol{\theta}_{exp}^{SM})$ though. The error associated with $\Delta S(m_h^{exp},\boldsymbol{\theta}_{exp}^{SM})$ can be computed by propagating the experimental errors of the experimental Higgs mass and the SM parameters to the entropies. In the Appendix~\ref{Appendix}, we compute the error of the new branching ratio, $\sigma_{BR_\chi}$, from the $1\sigma$ errors associated to all the SM parameters, taken from PDG~\cite{PhysRevD.98.030001}, that affect the computation of the Higgs entropy. From this computation, we found that
\begin{equation}
    4.1\% < BR_\chi(m_h^{exp},\boldsymbol{\theta}_{exp}^{SM},\boldsymbol{\theta}_\chi) < 11.3\%\; ,
    \label{eq:br_1sigma}
\end{equation}
with $\Delta S_{SM}(m_h^{exp},\boldsymbol{\theta}_{exp}^{SM}) = -0.00\pm 0.14$, 
so a positive $\Delta S_{SM}(m_h^{exp},\boldsymbol{\theta}_{exp}^{SM})$ is consistent with the experimental uncertainty pull in the entropy. Whenever a positive or null $\Delta S_{SM}(m_h^{exp},\boldsymbol{\theta}_{exp}^{SM})$ is no longer compatible with the data then a single new Higgs decay channel is not supported by the MEP inference.

%

This result is independent of the extension of the SM that predicts the new decay channel. It is true that $BR_\chi(m_h^{exp})\approx 1/(M+1)$ could have been predicted solely from an information theory argument -- the new probability distribution should be the less informative and more entropic guess. Our calculations confirm that intuition but, more importantly, it enables us to compute the error in $BR_\chi$ once we have Eqs.~\eqref{eq:sol} and \eqref{eq:deltaS}. Without this error estimate, the inference would not be complete or particularly useful. We now proceed to illustrate the MEP inference in new physics models.

\section{Applying MEP inference to new physics models}
\label{sec:typeI}

 The CMS measurement of the Higgs total width~\cite{Sirunyan:2019twz} can be used to put this universal MEP prediction to a first test. The measured value of $3.2^{+2.8}_{-2.2}$ MeV
 translates to $1.0\leq \Gamma^{exp}_H\leq 6.0$ MeV and if we suppose that the difference between the upper bound of $\Gamma_H^{exp}$ and the SM value of 4.07 MeV is due the contribution of a single new channel, then
 \begin{equation}
     BR_\chi\leq \frac{6.0-4.07}{6.0} = 32.2\%\;\; .
 \end{equation}
 So there is room for a new Higgs decay channel according to MEP for a while. In order to probe the $BR_\chi=7\%$, the precision of the Higgs width measurement should reach ${\cal O}(0.1\hbox{MeV})$, a task for the next runs of the LHC.
 
 The MEP inference on the branching ratio of the new channel, Eqs.~(\ref{eq:br_central}) and (\ref{eq:br_1sigma}), can now be readily used to predict the model parameters to a particular region of the parameters space. If further constraints exist, this might help to pin down the parameters of the model and test predictions in an accurate way. 
 
 The models that we might consider for MEP inference should fulfill some requirements: (1) there is only one new particle that the Higgs boson is allowed to decay into, the new spectrum might contain other particles but only $\chi$ should have a mass such that $m_\chi < m_h/2$; (2) the new spectrum should not contribute much to loop-induced decays of the Higgs into photons and gluons nor change the SM tree-level couplings, this is a very interesting possibility for what the MEP inference could be applied but it is beyond the scope of this work; (3) the model should allow to turn off couplings in order that only one new decay channel is accessible to the Higgs boson, if the spectrum contains more than one light particle, turning off some Higgs couplings in order that just one decay channel is possible is necessary for our results to apply. The following models might pass these conditions.
 
 \begin{figure}[!t]
     \centering
     \includegraphics[scale=0.45]{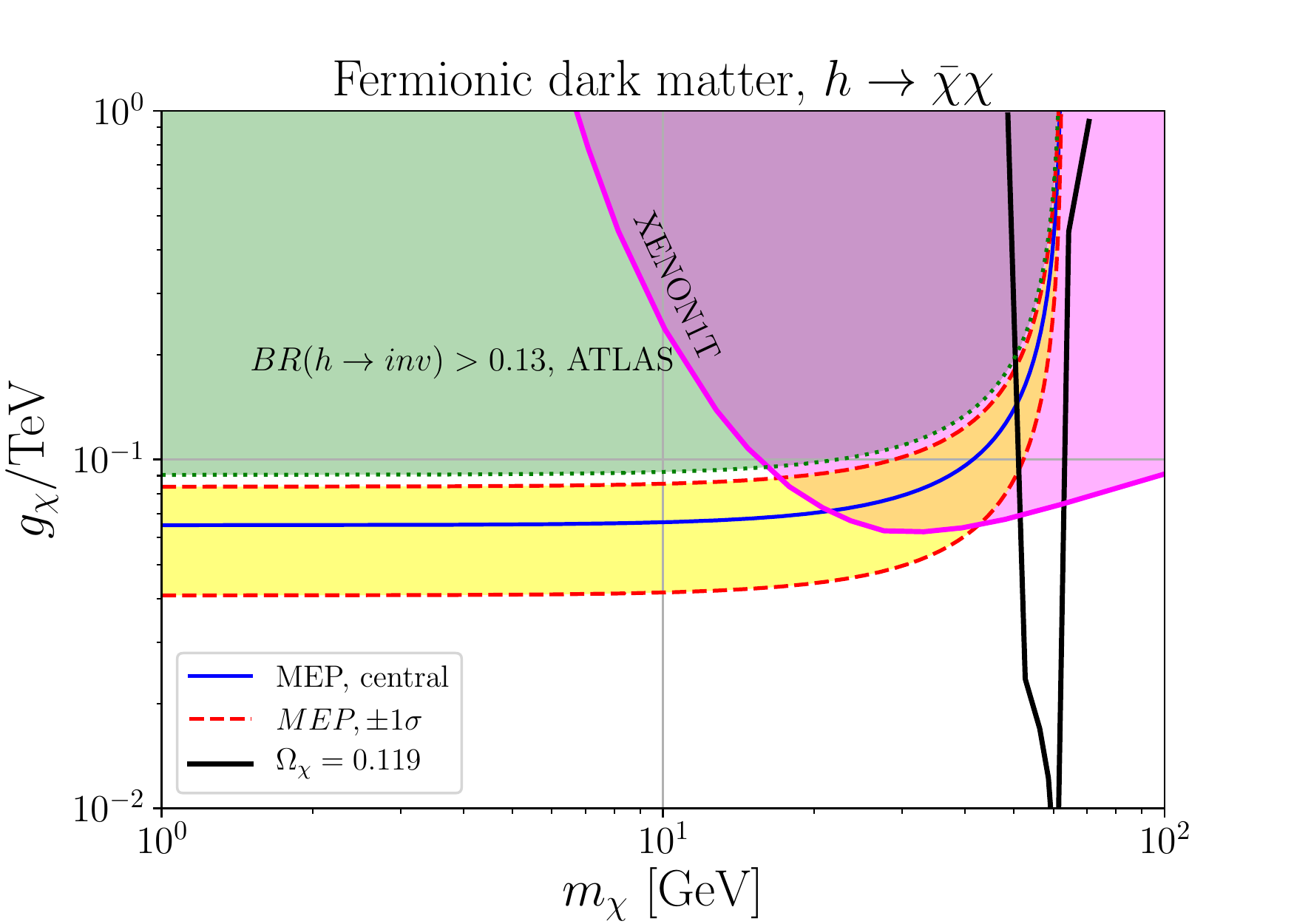}
     \includegraphics[scale=0.45]{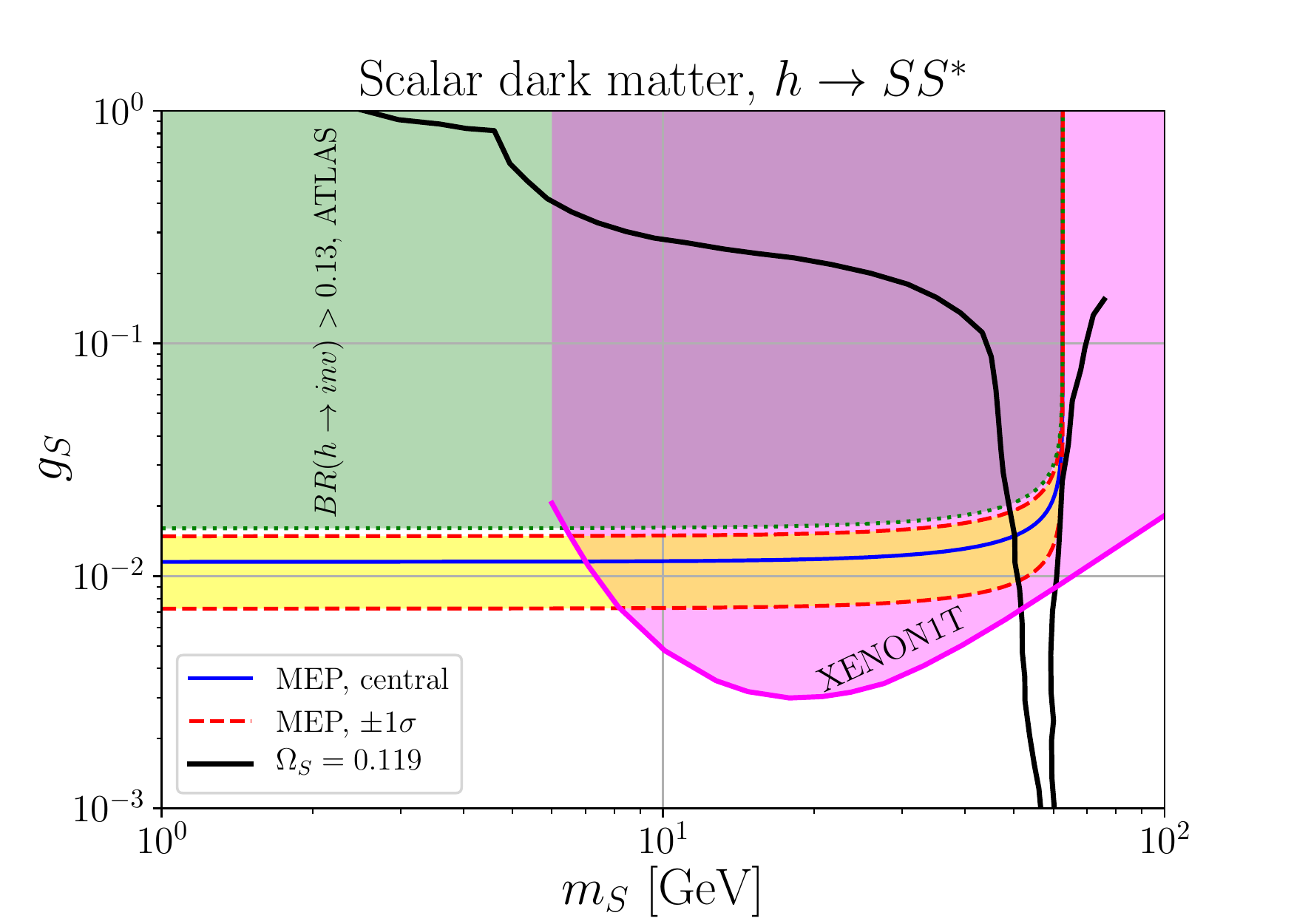}\\
     \includegraphics[scale=0.45]{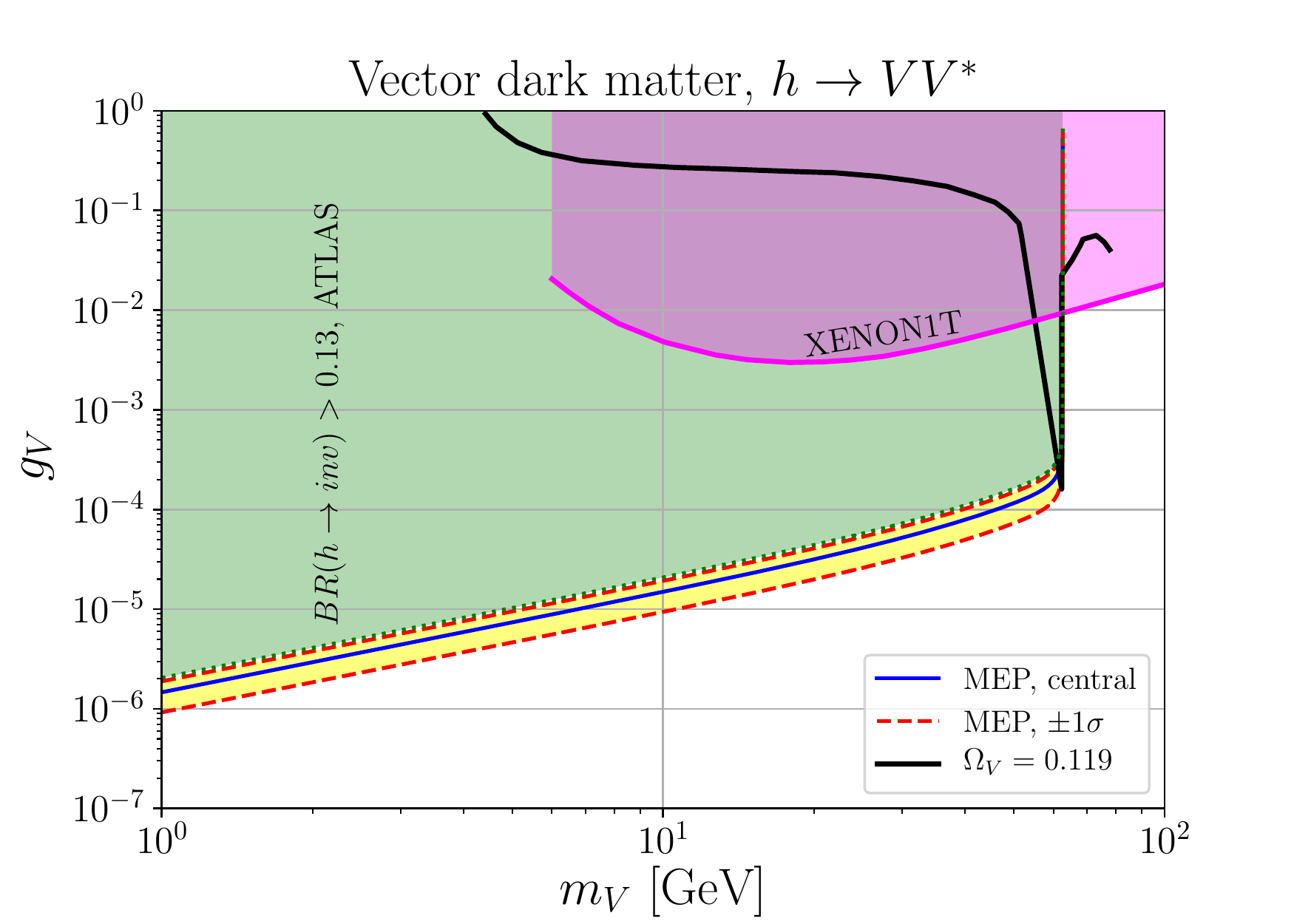}
     \includegraphics[scale=0.45]{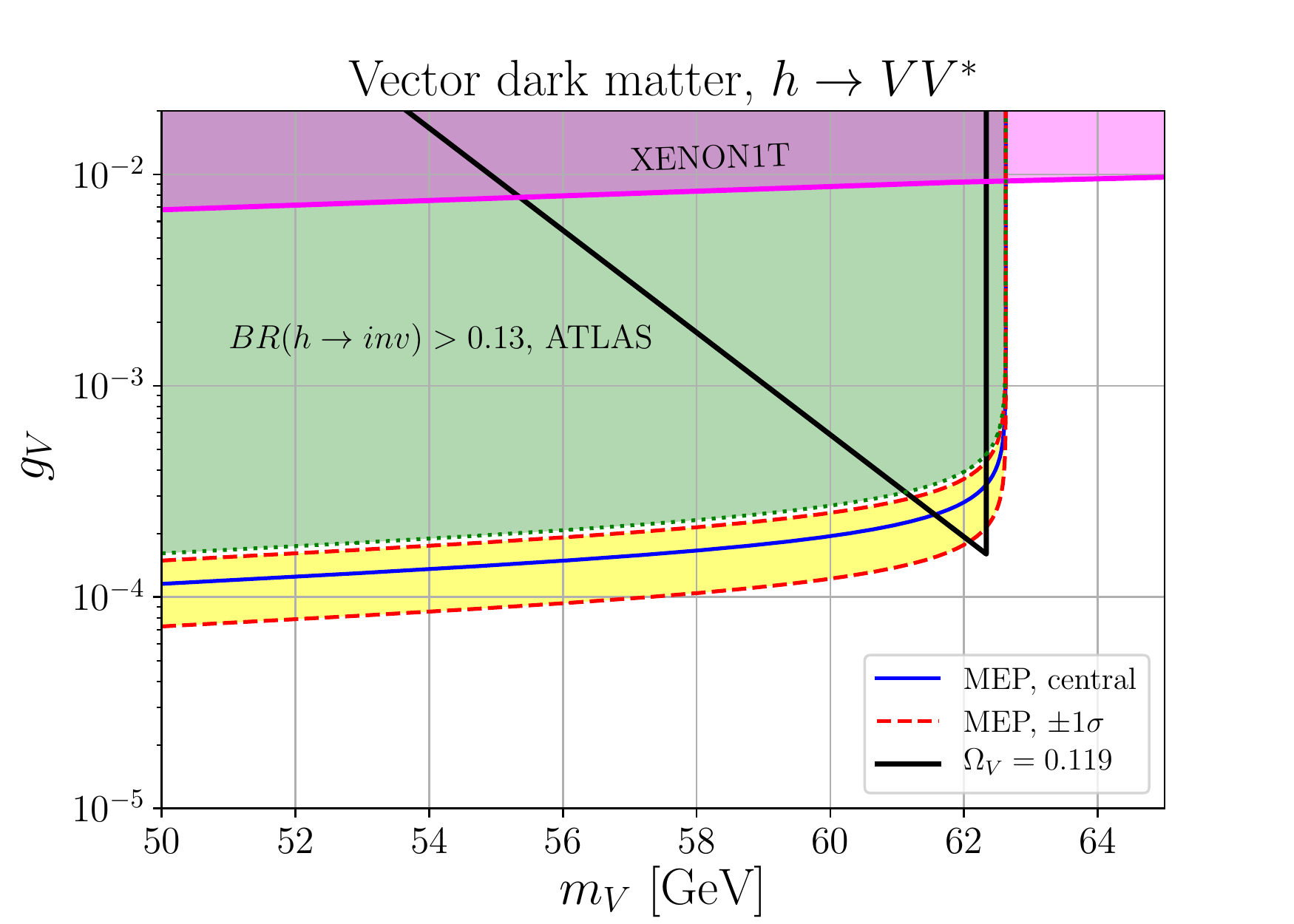}
     \caption{MEP applied to Higgs portal models for fermionic (upper left), scalar (upper right) and vector dark matter (lower plots). We show the central MEP prediction (solid blue lines) and its $1\sigma$ confidence region (yellow shaded regions) along with
     the LHC 95\% CL exclusion region of $h\to \hbox{invisible}$ (green shaded areas), the Xenon1T 90\% CL exclusion region (magenta shaded areas), and the points that satisfies the dark matter relic abundance as measured by Planck, $\Omega$, (the thick black lines) in the coupling {\it versus} DM mass plane. The lower right panel displays a zoom of the lower left panel in a region close to the dark matter production threshold, $50 < m_V < 65$ GeV.}
     \label{fig:portal}
 \end{figure}
 \subsection{Higgs-portal dark matter}
 \label{sub:dm}
 Suppose that a dark matter candidate interacts with the Higgs boson, $h$, through an effective operator in the cases where the dark matter is a fermion $\chi$, a real scalar $S$ or a vector $V$~\cite{Arcadi:2019lka}, and satisfy some symmetry that guarantees the DM stability,
 \begin{eqnarray}
     \Delta{\cal L}_F &=& -\frac{g_\chi v}{\sqrt{8}\Lambda} h \bar{\chi}\chi,\;\; \hbox{fermion} \\
     \Delta{\cal L}_S &=& -\frac{g_S v}{\sqrt{8}} h S^2,\;\; \hbox{real scalar} \\
     \Delta{\cal L}_V &=& -\frac{g_V v}{\sqrt{8}} h V_{\mu}V^\mu,\;\; \hbox{vector}
 \end{eqnarray}
 where $g_\chi$ is the effective coupling and $\Lambda$ an energy scale in the case of the fermionic dark matter; $g_S$ is the coupling to the scalar DM, and $g_V$ the coupling to the vector DM. The vacuum expectation value (vev) is denoted by $v$.  
 
 Three major additional constraints apply to this class of models -- (I) the relics abundance from Planck~\cite{Akrami:2018vks}, $\Omega_{DM}=0.1199\pm 0.0022$, (II) the LHC bound on invisible decays of the Higgs boson, $BR_{inv}<0.13$ at 95\% CL from ATLAS~\cite{ATLAS:2020cjb}, and (III) the 90\% CL direct detection limit on spin-independent scattering on nuclei from Xenon1T~\cite{Aprile:2017iyp}. The regions of the parameters space not excluded by these experiments should be compatible with the MEP inference. If the Higgs boson decays to some of these dark matter candidates, this new decay channel should, in conjunction with the SM channels, maximize the Higgs decays entropy within the experimental uncertainties as discussed in the previous section. 

 In Figure~\eqref{fig:portal}, we show the experimental constraints and the MEP inference of the coupling to the Higgs boson and the mass of the dark matter for the three Higgs-portal models. The blue solid lines correspond to the central MEP prediction, $BR_\chi = 7.1$\%. The yellow shaded regions display the parameters predicted by MEP within the uncertainties of the new branching ratio, that is it, is the region of the parameters space for which 4.1\% $< BR_\chi < $ 11.3\% according to Eq.~\eqref{eq:br_1sigma}.  The green shaded regions are the ones excluded by the LHC search for Higgs decaying invisibly. As we see, it is currently not strong enough to probe regions compatible with MEP but this might change in the next run of the LHC. The magenta regions are the portions of the parameters space excluded by Xenon1T. This experiment has not found any sign of direct dark matter scattering and has placed very strong bounds on dark matter models. Out of the three Higgs-portal models, we see that Xenon1T excludes, at 90\% CL, the region where MEP is compatible with relics abundance measured by Planck, the thick black lines, in the case of fermionic and scalar dark matter. However, the vector dark matter case is still viable, as we can see in the lower panels of Figure~\eqref{fig:portal}, and the MEP inference prefers the same region where the relics abundance falls in, this is the production threshold region where $m_V\sim m_h/2$ as we see in the zoomed region at the lower right panel. This region is likely to be probed by the LHC first.

 \subsection{Dark $Z$ boson model}
 A massive dark gauge boson, $Z_D$, coupled to the Higgs boson might lead to a new Higgs decay depending on the dark $Z$ mass, $m_{Z_D}$. A new gauge boson can couple to Higgs bosons through CP-even dimension-3, $h Z_\mu Z_D^\mu$ ($h A_\mu Z_D^\mu$ is prohibited by gauge invariance), and dimension-5 operators, $hX_{\mu\nu}Z^{\mu\nu}_D$~\cite{Davoudiasl:2013aya}. These interactions arise from kinetic mixing between the SM $X=Z,A$ and the dark boson $Z_D$ 
 \begin{equation}
     \Delta {\cal L}_{Z_D} = \sum_{X=A,Z} \left[C_X h X_{\mu\nu}Z^{\mu\nu}_D + \frac{\tilde{C}_X}{2} \varepsilon_{\mu\nu\rho\sigma} hX^{\mu\nu}Z^{\rho\sigma}_D\right] +\frac{\epsilon}{2\cos\theta_W} (-\sin\theta_W Z_{\mu\nu}+\cos\theta_W F_{\mu\nu})Z^{\mu\nu}_D ,
 \end{equation}
where $\tilde{C}_X$ denote CP-odd interactions in the case of dimension-5 operator, $X_{\mu\nu}=\partial_\mu X_\nu-\partial_\nu X_\mu$, where $X_{\mu\nu}=Z_{\mu\nu},F_{\mu\nu}$ are the field strength tensors of the SM $Z$ boson and the photon, respectively, and $Z^{\mu\nu}_D=\partial^\mu Z_D^\nu-\partial^\nu Z_D^\mu$ is the field strength tensor of the $Z_D$ dark boson; $\theta_W$ is the Weinberg angle.
 
 The Higgs boson can decay into $ZZ_D$, $Z_DZ_D$ and $\gamma Z_D$, but if $m_h/2 < M_{Z_D} < m_h$ then only the $\gamma Z_D$ is open. Let us call the effective $\gamma$-$Z_D$-$h$ interaction as $\kappa_\gamma=C_A+\tilde{C}_A/2$. In Ref.~\cite{Sirunyan:2019xst}, the CMS Collaboration has performed a search for the $h\to \gamma+$ invisible in the $Zh$ channel after 137 fb$^{-1}$ of data at the 13 TeV LHC. The study placed a 95\% CL upper limit on the branching ratio of $BR(h\to \gamma+\hbox{invisible})<4.6$\%. If this is the only new decay channel of the Higgs boson in a hypothetical new physics model, than the LHC has almost entirely excluded this possibility once it is barely compatible with the MEP prediction. 
\begin{figure}[!t]
     \centering
     \includegraphics[scale=0.46]{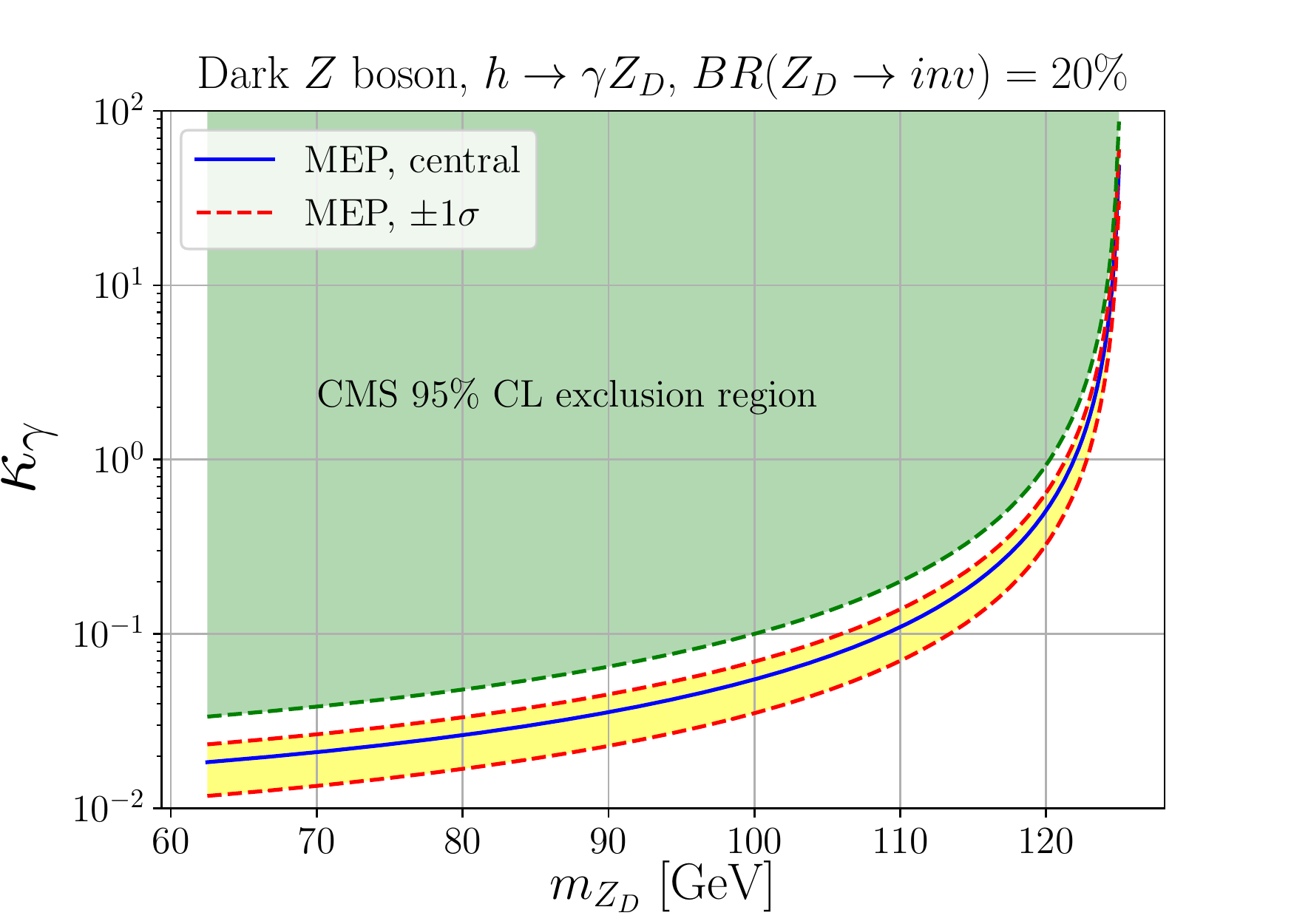}
     \includegraphics[scale=0.46]{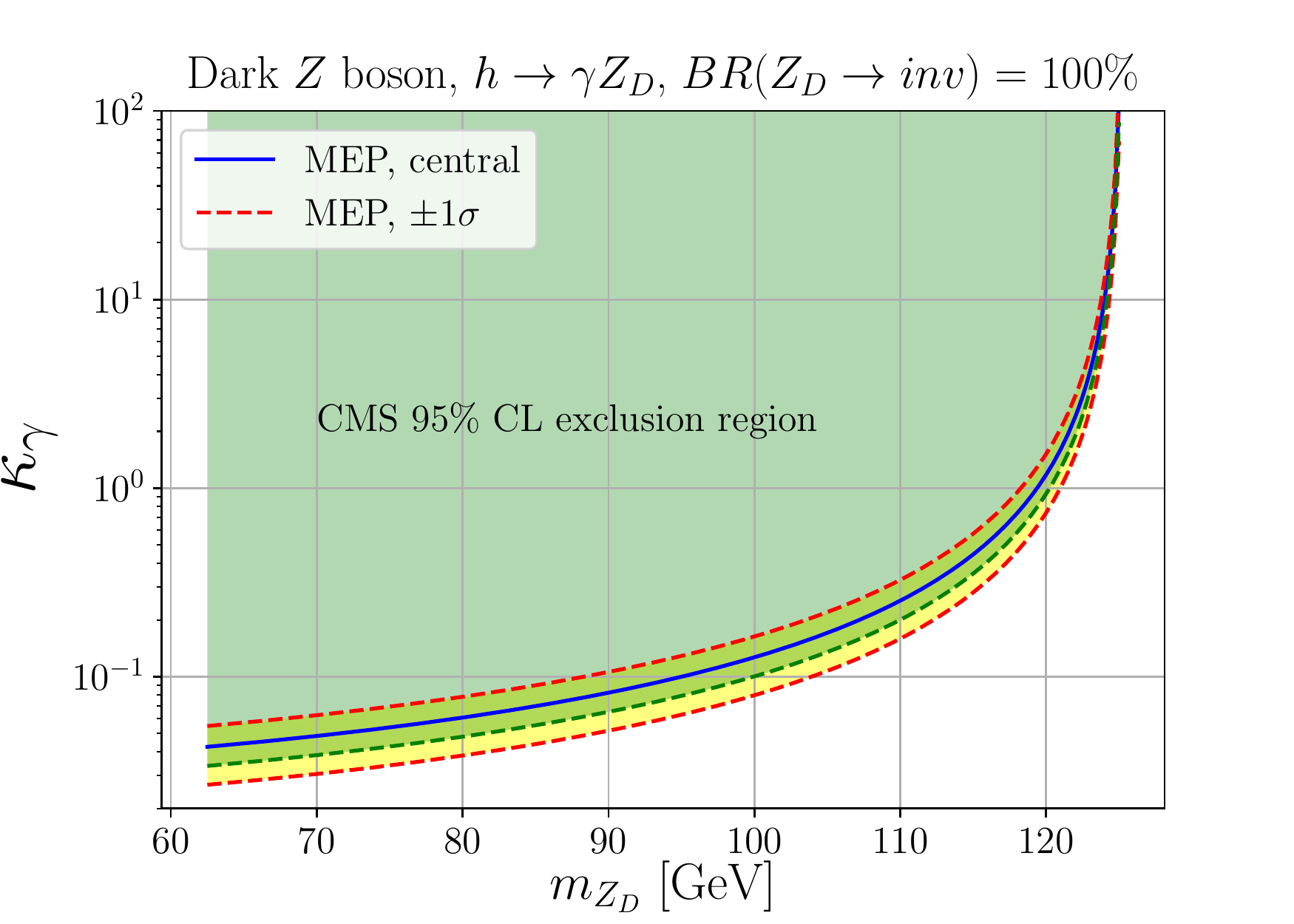}
     \caption{The MEP inference of the dark gauge boson model described in the text. We show two scenarios, at left the case where $Z_D$ decays invisibly 20\% of the time, while the case where it always decays invisibly at right in the plane of effective coupling {\it versus} $Z_D$ mass. The green shaded area is excluded by the LHC at 95\% CL. The blue line and the yellow band depict the parameters of the model compatible with MEP including the propagated errors of the SM parameters.}
     \label{fig:darkZ}
 \end{figure}

The mixing of the SM $Z$ boson and the dark gauge boson $Z_D$ permits that the new gauge boson decays to SM fermions. If the $Z_D$ dark gauge boson decays to neutrinos, than the CMS constraint hits the model and can be used to put bounds on its parameters. In Figure~\eqref{fig:darkZ}, left panel, we show the MEP inference in the $m_{Z_D}\times \kappa_\gamma$ plane in the case where $BR(Z_D\to \hbox{invisible})=20$\%, and at the right panel, the case where $BR(Z_D\to \hbox{invisible})=100$\%. Only if the $BR(Z_D\to \hbox{invisible})$ is large, the current experimental constraints can probe the region predicted by MEP. In the next run of the LHC, the entire region is likely to be probed though.

 \subsection{Lepton flavor violation models}
A single lepton flavor violating (LFV) coupling to the Higgs boson can be modeled as~\cite{Harnik:2012pb}
\begin{equation}
     \Delta{\cal L}_{LFV} = -\Upsilon_{ij}\bar{\ell}^i_L \ell^j_R h + \hbox{h.c.}
 \end{equation}
 where $\Upsilon_{ij}$ is the Yukawa coupling between a fermion pair of different flavors, $i\neq j$. In new physics models, the Yukawa matrix $\Upsilon$ need not to be diagonal.
 
 A modest excess of $h\to \mu\tau$ was observed by the CMS back to 2015~\cite{Khachatryan:2015kon} but not confirmed by ATLAS~\cite{Aad:2015gha}.
 For a Higgs mass much heavier than the fermion masses, the new branching fraction into leptons of different flavor is proportional to $(\Upsilon_{ij}^2+\Upsilon_{ji}^2)^{1/2}$, and effectively, the region compatible with MEP lies between two concentric circles as depicted at the left panel of Figure~\eqref{fig:alp-lfv} where we show a LFV model example for Higgs decaying into a tau lepton-muon pair. 
 
 The region between the dashed red circles is allowed by MEP, but the green region was later excluded by the CMS searches for his type of Higgs decay at 95\% CL~\cite{Sirunyan:2017xzt}: $(\Upsilon_{\mu\tau}^2+\Upsilon_{\tau\mu}^2)^{1/2}<1.43\times 10^{-3}$. The experimental search excludes the region where this new decay channel would be compatible to the MEP inference. Therefore, based on the MEP prediction, this type of lepton flavor violation interaction of the Higgs boson should never be observed alone. If it is observed in future searches but no other LFV decays, then MEP applied to Higgs decays is falsified. 
 
 Many models can be built to predict LFV couplings of the Higgs boson. It is out of the scope of our work to revise and discuss how each one of these models could be affected by this inference. We, however, point out that the results of this work apply to one channel at once. In the case where many channel are taken into account at the same time, the approach should be that one described in Ref.~\cite{Alves:2017ljt}. In the case of LFV models, it is always possible to adjust the Yukawa couplings in order to have just one new decay channel. The plausibility of this scenario should be discussed within each model framework.
 
 \subsection{Higgs-Axion Like Particle effective coupling models}
The leading Higgs-Axion Like particle (ALP) interaction arises from the following dimension-6 operator~\cite{Bauer:2017ris}
\begin{equation}
     \Delta{\cal L}_{ha} = \frac{C_{ah} v}{\Lambda^2} (\partial_\mu a)(\partial^\mu a) h,
 \end{equation}
 where $a$ denotes the ALP field of mass $m_a$ and $C_{ah}/\Lambda$ the effective Higgs-ALP coupling.
 \begin{figure}[!t]
     \centering
     \includegraphics[scale=0.4]{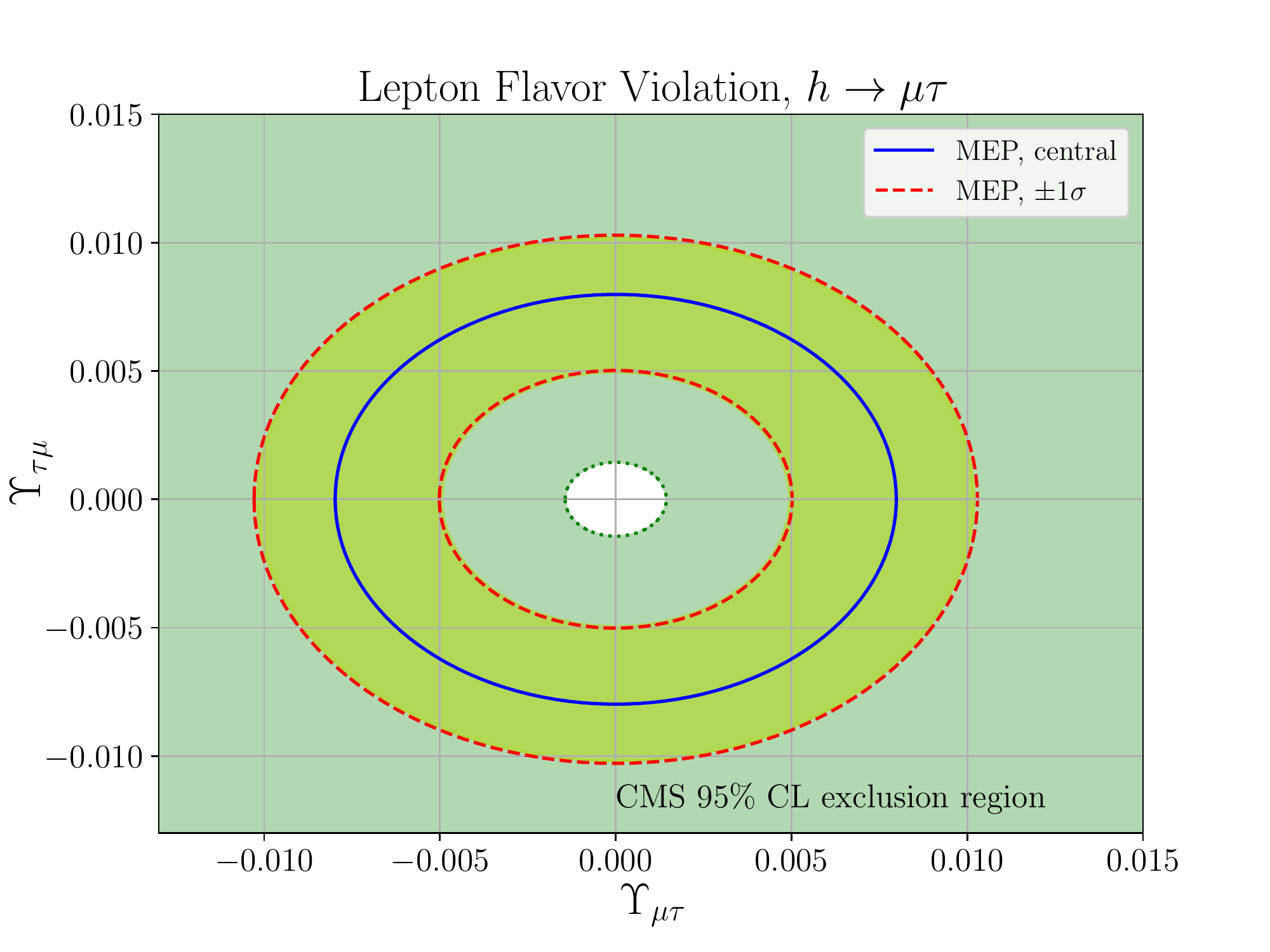}
      \includegraphics[scale=0.48]{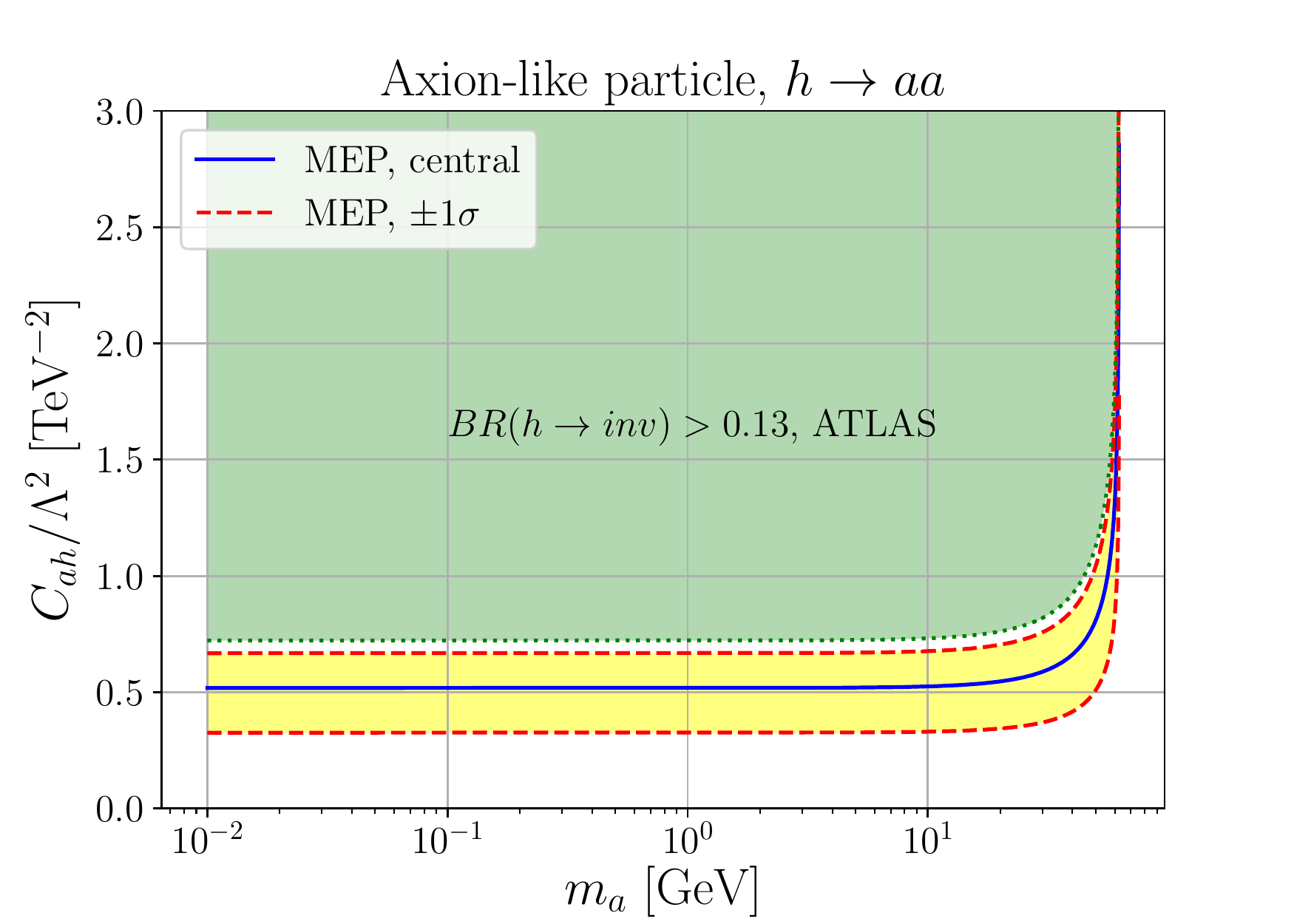}
     \caption{At the left panel we show the MEP inference of the LFV model where the Higgs boson interacts with a muon-tau lepton pair. At the right panel, we show the inference of the parameters of a model where the Higgs interacts with an axion-like particle and the ALP decays invisibly or outside the detectors. In both panels, the green shaded region is excluded by LHC searches at 95\% CL~\cite{Sirunyan:2017xzt} at left, and~\cite{ATLAS:2020cjb} at right. The blue line and the yellow band depict the parameters of the model compatible with MEP including the propagated errors of the SM parameters.}
     \label{fig:alp-lfv}
 \end{figure}
 
 In Figure~\eqref{fig:alp-lfv}, at the right panel, we show the region of the mass and coupling, $m_a \times C_{ah}/\Lambda^2$, plane. The MEP constraint is pretty insensitive to the ALP mass up to masses close to $m_h/2$. If the ALP couplings to photons and other particles are suppressed in such a way that the ALP is long-lived enough to decay outside the detectors and also supposing that 100\% of all ALPs lead to an invisible signature in Higgs decays, we can impose the LHC bound on Higgs invisible decays to this case as well. If the ALP is a dark matter candidate, many other constraints apply, just like Higgs portal to scalar DM discussed in Section~\ref{sub:dm}. The region excluded by the LHC searches of $h\to \hbox{invisible}$ is shown in the green shaded area at the right panel of Figure~\eqref{fig:alp-lfv}, right panel. Again, it is very likely that the next run of the LHC will probe the region where MEP predicts that such Higgs decay occur in this ALP model. 
 
 Another example of a Higgs boson decaying to scalars can be found in Ref.~\cite{Robens:2019kga}. In that work, the branching ratio to new scalars after applying theoretical and experimental constraints was found to be up to 7--8\%, which might be in agreement to what MEP predicts.

\section{Conclusions}
\label{sec:conclusions}
 The Maximum Entropy Principle is a powerful tool for statistical inference in several branches of science. In particle physics, MEP found extraordinary success in predicting the Higgs boson mass by computing the entropy of the Higgs boson decay probabilities. It is therefore natural to postulate that any new Higgs decay channel should respect the principle and then investigate the consequences. 
 
 We found that requiring that a single extra decay channel added to the SM channels should still maximize the entropy close to the experimentally observed mass leads to an universal prediction -- the new decay channel must account for around 7\% of all the Higgs decays, irrespective of the type of new physics. This result can be easily understood in information theory basis, it is just the maximally uninformed guess that could be done for a single new channel. We, however, worked out an estimate of the uncertainty in this new branching ratio from the solution to a differential equation governing the new branching fraction in terms of the Higgs mass. With the current uncertainties in the SM parameters, the error in the new branching is $\pm 3$\% approximately.
 
 The prescription to predict the parameters of an extended model is now straightforward using MEP, just require that the new branching ratio $BR_\chi$ satisfies
 \begin{equation}
    4.1\% < BR_\chi(m_h^{exp},\boldsymbol{\theta}_{SM}^{exp},\boldsymbol{\theta}_\chi) < 11.3\%\; ,
 \end{equation}
 where $m_h^{exp},\boldsymbol{\theta}_{SM}^{exp}$, and $\boldsymbol{\theta}_\chi$ are the measured Higgs mass and SM parameters, and the parameters of the new model, respectively. The range predicted by MEP is still permitted by the LHC measurement of the Higgs width~\cite{Sirunyan:2019twz}. 
 
 We propose tests to the MEP inference to four classes of models predicting a new decay channel of the Higgs boson. In the cases of Higgs-ALP and Higgs-dark gauge boson interactions, we determined the region of the coupling {\it versus} mass plane that is compatible with MEP. Supposing that the ALP decays outside the detectors and the dark gauge gauge boson decays invisibly, we found that the current LHC bound on the Higgs decay to invisible states is not strong enough to probe the ALP model that we considered, but it might probe the parameters of a dark gauge boson model predicted by MEP. In the case of Higgs portal models and lepton flavor violation models, strong experimental constraints are at disposal. The LHC search for $h\to \mu\tau$, for example, excluded the entire region of the model parameters space that fulfills MEP. If the principle is correct, no such LFV decay of the Higgs boson exists. Higgs portal models with fermionic, scalar and vector dark matter are constrained by many experiments. We found that, currently, only the vector dark matter still presents parameters compatible with MEP and not excluded by the LHC, Planck or Xenon1T data. These predictions should be confronted to more experimental searches as soon as they are available.
 
 The MEP predictions to BSM scenarios can establish it as an useful inference tool in Higgs physics, helping to spot regions of the parameters space of a model. If it works, the benefits for particle phenomenology will be many. On the other hand, if MEP pass further tests, it might also shed light on the mechanism that adjusts the Higgs boson mass or even other parameters of the theory. 
 Maybe MEP is pointing to a physical entropic mechanism that took place at some early stage of the evolution of the universe. We invite the community to think seriously about this intriguing possibility.
 
\vspace{0.3 cm}
 
{\bf{Acknowledgments:}}
The authors thank Farinaldo Queiroz for the help with the Xenon1T data. A. Alves, A. G. Dias and R. da Silva thank Conselho Nacional de Desenvolvimento Cient\'{\i}fico (CNPq) 
for its financial support, under grants 307265/2017-0, 311236/2018-9, and 424052/2018-0, respectively.   
  
\appendix
\section{Uncertainty in the new branching ratio of the Higgs boson}
\label{Appendix}

The uncertainty in the new Higgs branching ratio $\sigma_{BR_{\chi }}$ can be obtained from error propagation.
From Eqs.~\eqref{eq:sol} and \eqref{eq:deltaS}, we see that this uncertainty depends on the uncertainties of
all the SM parameters which affect the SM branching ratios.

With the uncertainty in $BR_{\chi }$ in hands,
we are able to obtain the uncertainty in the coupling and the mass of the new particle. 
We therefore start computing the uncertainty in the $\Delta
S=S_{SM}\left(m_h^{exp}\right)-S_{SM}\left(m_h^{sm}\right)$, which is calculated
according to 
\begin{equation}
\sigma _{\Delta S}=\sqrt{\sigma _{S_{m_h^{exp}}}^{2}+\sigma
_{S_{m_h^{sm}}}^{2}}\;\; ,  
\label{Eq:Propag_1}
\end{equation}%
where $\sigma _{S_{m_h^{exp}}}$ and $\sigma
_{S_{m_h^{sm}}}$ denote the uncertainties of $S_{SM}$ considering the experimental mass 
 and mass inferred with MEP, respectively. 

The entropy $S_{SM}$ is a function of $m_h$ and $p$ SM parameters $%
\boldsymbol{\theta}_{SM}=(z_{1}$, $z_{2},...,z_{p})$. All these parameters have experimental
uncertainties $\sigma _{z_{1}}$, $\sigma _{z_{2}},...,\sigma _{z_{p}}$.
Then, to compute the uncertainty in $m_h^{sm}$, we estimate
the probability density function of the Higgs mass as determined by MEP, $P(m_h^{sm})$, 
by marginalizing over all the $p$ SM parameters. 
This is done by generating a random sample of SM parameters 
$\boldsymbol{\theta}_{SM}^k=(z_{1}^{(k)},...,z_{p}^{(k)}), k=1,\cdots, N$ drawn 
from independent Gaussian distributions for each parameter with mean $\overline{z}$ and standard deviation $\sigma_{z}$,
 and computing the Higgs mass inference for each one of the $N$ samples
$\{m_{h,k}^{sm}\}_{k=1}^{N}=\{m_{h,k}^{sm}=\underset{m_h}{\mathrm{argmax}}\; S_{SM}(m_h,\boldsymbol{\theta}^{k}_{SM}),k=1,\cdots, N\}$.   
Now, considering all $N$ $p$-uples, we have a
sample $\{m_{h,k}^{sm}\}_{k=1}^{N}$ so we
can estimate the average $\overline{m}_h^{sm}=(1/N)%
\sum_{k=1}^{N}m_{h,k}^{sm}$ and the uncertainty $\sigma
_{}$ as the standard deviation of average, i.e., $%
\sigma _{m_h^{sm}}=s_{m_h^{sm}}/\sqrt{N}$, where $%
s_{m_h^{sm}}=\sqrt{\frac{1}{N-1}\sum%
\limits_{k=1}^{N}(m_{h,k}^{sm}-\overline{m}_h^{sm})^{2}}$.

On the other hand, we have an experimental estimate for $m_h^{exp}$,
described by the confidence interval $\overline{m}_h^{exp}\pm \sigma
_{m_h^{exp}}$. So, we can calculate $\sigma _{S_{m_h^{exp}}}$ and $%
\sigma _{S_{m_h^{sm}}}$ propagating uncertainties and
using numerical estimates of derivatives. Considering that $%
S_{SM}=S_{SM}(m_h,z_{1},...,z_{p})$, we have that
\begin{eqnarray}
\sigma _{S_{m_h^{exp}}}^{2} &=&\left( \left. \frac{\partial
S_{SM}(m_h,z_{1},\cdots,z_{p})}{\partial m_h}\right\vert _{m_h=\overline{m%
}_{h}^{exp},z_{i}=\overline{z}_{i},i=1,\cdots,p}\right) ^{2}\sigma
_{_{m_h^{exp}}}^{2}  \label{Eq:sig_exp_1} \\
&&+\sum\limits_{i=1}^{p}\left( \left. \frac{\partial S(m_h,z_{1},\cdots,z_{p})%
}{\partial z_{i}}\right\vert _{m_h=\overline{m}_h^{exp},z_{i}=%
\overline{z}_{i},i=1,\cdots,p}\right) ^{2}\sigma _{z_{i}}^{2}  \notag
\end{eqnarray}

In first aproximation, following the prescription used in \cite{Brusamarello2008}, whe have 
\begin{equation}
\begin{array}{lll}
\left. \frac{\partial S_{SM}(m_h,z_{1},\cdots,z_{p})}{\partial m_h}%
\right\vert _{m_h=\overline{m}_h^{exp},z_{i}=\overline{z}%
_{i},i=1,\cdots,p} & = & \frac{S_{SM}(\overline{m}_h^{exp}+\sigma _{_{%
\overline{m}_h^{exp}}},\overline{z}_{1},...,\overline{z}_{p})-S_{SM}(%
\overline{m}_h^{exp}-\sigma _{_{\overline{m}_h^{exp}}},\overline{z}_{1},\cdots,%
\overline{z}_{p})}{2\sigma _{_{\overline{m}_h^{exp}}}}+O(\sigma _{_{%
\overline{m}_h^{exp}}}^{2}) \\ 
&  &  \\ 
& \approx & \frac{S_{SM}(\overline{m}_h^{exp}+\sigma _{_{\overline{m}_{h}^{exp}}},\overline{z}_{1},\cdots,\overline{z}_{p})
-S_{SM}(\overline{m}_{h}^{exp}-\sigma _{\overline{m}_h^{exp}},\overline{z}_{1},...,%
\overline{z}_{p})}{2\sigma _{_{\overline{m}_h^{exp}}}}%
\end{array}
\label{Eq:aproximation}
\end{equation}

Substituting the Eq.~\eqref{Eq:aproximation} in Eq.~\eqref{Eq:sig_exp_1},
and extending this approximation to the other parameters, we have

\begin{equation}
\begin{array}{lll}
\sigma _{S_{m_h^{exp}}}^{2} & \approx & \frac{1}{4}\left[ S_{SM}(%
\overline{m}_h^{exp}+\sigma _{\overline{m}_h^{exp}},\overline{z}%
_{1},\cdots,\overline{z}_{p})-S_{SM}(\overline{m}_h^{exp}-\sigma _{_{%
\overline{m}_h^{exp}}},\overline{z}_{1},\cdots,\overline{z}_{p})\right] ^{2}
\\ 
&  &  \\ 
&  & +\frac{1}{4}\sum\limits_{i=1}^{p}\left[ S_{SM}(\overline{m}_h^{exp},%
\overline{z}_{1},\cdots,\overline{z}_{i-1},\overline{z}_{i}+\sigma _{z_{i}},%
\overline{z}_{i+1},\cdots,\overline{z}_{p})-S_{SM}(\overline{m}_h^{exp},%
\overline{z}_{1},\cdots,\overline{z}_{i-1},\overline{z}_{i}-\sigma _{z_{i}},%
\overline{z}_{i+1},\cdots,\overline{z}_{p})\right] ^{2}%
\end{array}%
\end{equation}

And similarly
\begin{equation}
\begin{array}{lll}
\sigma _{_{S_{m_h^{sm}}}}^{2} & \approx & \frac{1}{4}\left[ S_{SM}(%
\overline{m}_h^{sm}+\sigma _{},\overline{z}_{1},\cdots,%
\overline{z}_{p})-S_{SM}(\overline{m}_h^{sm}-\sigma _{},%
\overline{z}_{1},\cdots,\overline{z}_{p})\right] ^{2} \\ 
&  &  \\ 
&  & +\frac{1}{4}\sum\limits_{i=1}^{p}\left[ S_{SM}(\overline{m}_h^{sm},%
\overline{z}_{1},\cdots,\overline{z}_{i-1},\overline{z}_{i}+\sigma _{z_{i}},%
\overline{z}_{i+1},\cdots,\overline{z}_{p})-S_{SM}(\overline{m}_h^{sm},%
\overline{z}_{1},\cdots,\overline{z}_{i-1},\overline{z}_{i}-\sigma _{z_{i}},%
\overline{z}_{i+1},\cdots,\overline{z}_{p})\right] ^{2}%
\end{array}%
\end{equation}

Given $\Delta S$, the new branching ratio is obtained by solving the transcendental Eq.~\eqref{eq:sol}.
Knowing the uncertainty $\sigma _{\Delta S}$ of $\Delta S$,
which is calculated according to Eq.~\eqref{Eq:Propag_1}, we can solve for $BR_{\chi }(\overline{\Delta S}+\sigma _{\Delta S})$ and $%
BR_{\chi }(\overline{\Delta S}-\sigma _{\Delta S})$, where

\begin{figure}[!t]
     \centering
     \includegraphics[scale=0.5]{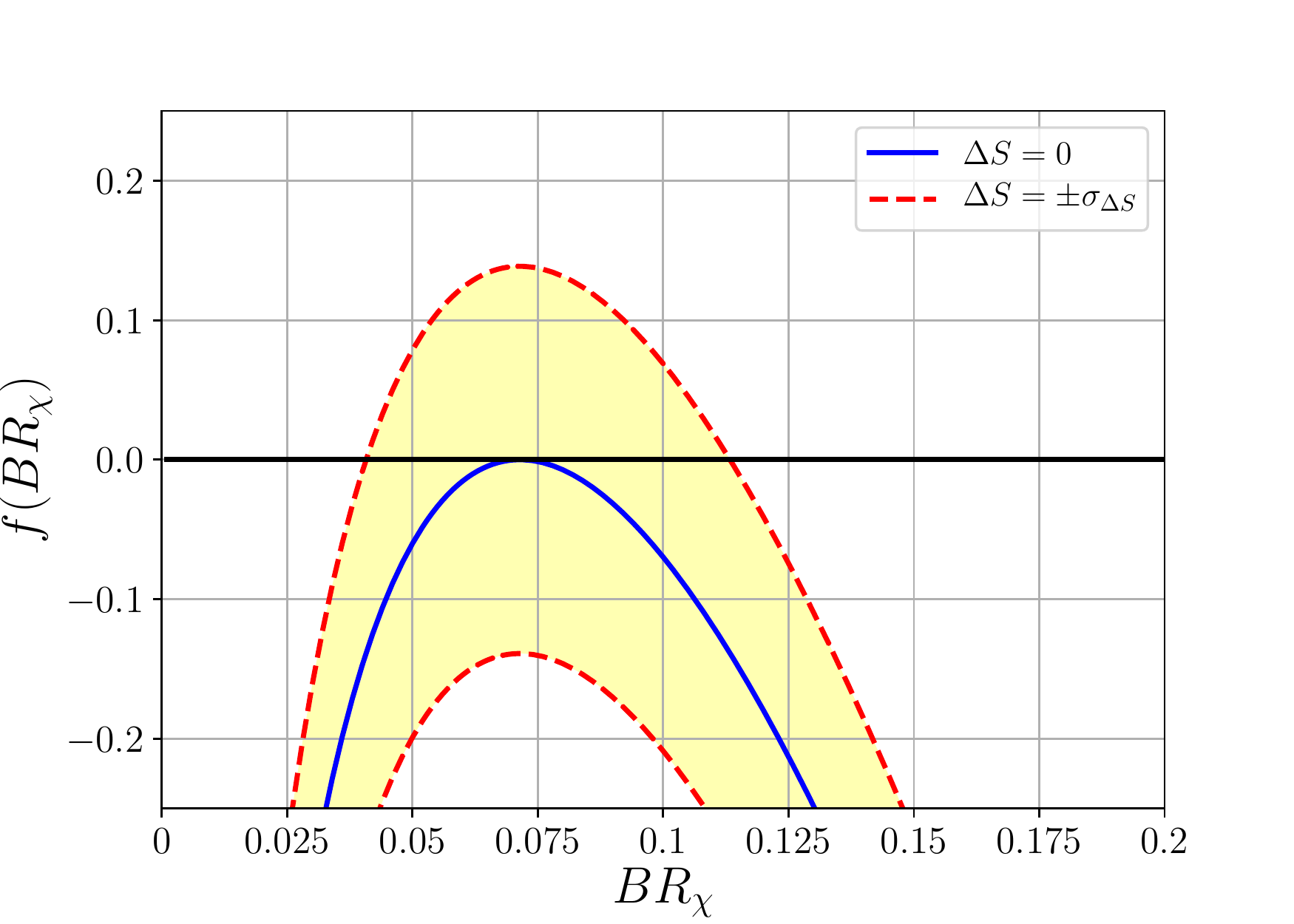}
     \caption{The logarithm of Eq.~\eqref{eq:sol} as a function of the new Higgs branching ratio, $BR_\chi$. The blue solid line shows the case where $\Delta S=0$ and only one real solution exists, $BR_\chi(m_h^{sm})=7.1$\%. The dashed red lines show the cases where $\Delta S=\pm \sigma_{\Delta S}$.}
     \label{fig:sigmaBR}
 \end{figure}

\begin{equation}
\overline{\Delta S}=S_{SM}(\overline{m}_h^{exp},\overline{z}_{1},\cdots,%
\overline{z}_{p})-S_{SM}(\overline{m}_h^{sm},\overline{z}_{1},\cdots,%
\overline{z}_{p})
\end{equation}
and the we have an estimate for $\sigma_{BR_{\chi }}$ given by 
\begin{equation}
\begin{array}{lll}
\sigma _{BR_{\chi }} & = & \sqrt{\left( \frac{\partial BR_{\chi }}{\partial
\Delta S}\right) ^{2}\sigma _{\Delta S}^{2}} \\ 
&  &  \\ 
& \approx  & \frac{1}{2}\left\vert \max(1,BR_{\chi }(\overline{\Delta S}+\sigma
_{\Delta S}))-\min(0,BR_{\chi }(\overline{\Delta S}-\sigma _{\Delta S}))\right\vert \; .
\end{array}%
\end{equation}

 In the equation above, we have to be careful that the branching ratio is bounded in the $[0,1]$ interval. In practice, in our case, because Eq.~\eqref{eq:sol} has no real solution if $\Delta S<0$, $\sigma _{BR_{\chi }} \approx  \frac{1}{2} \max(1,BR_{\chi }(\overline{\Delta S}+\sigma
_{\Delta S}))$. In Figure~\eqref{fig:sigmaBR}, we show 
\begin{equation}
    f(BR_\chi)=\ln BR_\chi+M\ln(1-BR_\chi)-\ln\frac{M^M}{(M+1)^{(M+1)}}+\Delta S_{SM}(m_h,\boldsymbol{\theta}_{SM})\; ,
    \label{eq:fbr}
\end{equation}
the logarithm of Eq.~\eqref{eq:sol} as a function of $BR_\chi$ for fixed $m_h$ and $\boldsymbol{\theta}_{SM}$. The solid blue line shows $f(BR_\chi)$ for $\Delta S(m_h=m_h^{sm},\boldsymbol{\theta}_{SM}^{exp})=0$. In this case, we have the central prediction of MEP as the only solution to Eq.~\eqref{eq:sol}, $BR_\chi(m_h^{sm})=7.1$\%. The yellow shaded region between the upper dashed line, where $\Delta S=+\sigma_{\Delta S}$, and the lower dashed line where $\Delta S=-\sigma_{\Delta S}$, denotes the variation in the inferred branching ratio due the uncertainties in the experimental Higgs mass and the other SM  parameters. Note that if $\Delta S < 0$, there is no real solution as we anticipated. For $\Delta S=+\sigma_{\Delta S}$, however, we find two solutions to $BR_\chi$: $4.1$\% and $11.3$\%.

\bibliography{main_rev3}
\end{document}